\documentclass{aa}  

\usepackage{graphicx}
\usepackage{amsmath} 
\usepackage[utf8]{inputenc}
\usepackage{hyperref}

\newcommand{\upsub}[1]{\sb{\mathrm{#1}}}
\begingroup\lccode`~=`_\lowercase{\endgroup\let~\upsub}

\AtBeginDocument{%
  \catcode`_=12
  \mathcode`_="8000
}

\usepackage{txfonts}
\usepackage{natbib}
\usepackage{footmisc}
\bibpunct{(}{)}{,}{a}{}{,}
\newcommand{\kmps}{km\,s$^{-1}$}

\newcommand{\dzcob}{$^{12}$CO }

\newcommand{\tzcob}{$^{13}$CO }

\newcommand{\nddpb}{N$_2$D$^+$ }

\newcommand{\juzb}{(\emph{J}:1 -- 0) }

\usepackage{color}
\usepackage[normalem]{ulem}

\begin{document}

   \title{Radio telescope total power mode: improving observation efficiency\thanks{Based on observations carried out with the Green Bank Telescope and the IRAM 30-m. The 
   Green Bank 
   Observatory is a facility 
   of the National 
   Science Foundation operated under cooperative agreement by Associated Universities, Inc. IRAM is
      supported by INSU/CNRS (France), MPG (Germany), and IGN (Spain).}}


   \author{Laurent Pagani           \inst{1}
       \and David Frayer\inst{2}
       \and Bruno Pagani \inst{3}
       \and Charl\`ene Lef\`evre \inst{4}
         }

   \offprints{L.Pagani}

 \institute{ 
 LERMA \& UMR8112 du CNRS, Observatoire de Paris, PSL University, Sorbonne Universités,  CNRS, F-75014 Paris, France\\
\email{laurent.pagani@obspm.fr}
 \and
 Green Bank Observatory, P.O. Box 2, Green Bank, WV 24944, USA
 \and
 Lyc\'ee Jean-Baptiste Corot, F-91605 Savigny-sur-Orge, France
 \and
 Institut de Radioastronomie Millim\'etrique (IRAM), 300 rue de la Piscine, 38400 Saint-Martin d’He\`eres, France
           }

   \date{Received July 20, 2020; accepted September 19, 2020}

 
  \abstract 
   {} 
   {{Radio }observing efficiency can be improved by calibrating and reducing the observations in total power mode
   rather than in frequency, beam, or position-switching modes.} 
   {{We selected a sample of spectra obtained from the Institut de Radio-Astronomie Millim\'etrique (IRAM) 30-m telescope
    and the Green Bank Telescope (GBT) to test the feasibility of  the method. 
        Given that modern front-end amplifiers for the GBT and direct Local Oscillator injection for the 30 m telescope provide smooth pass bands
        that are a few tens of megahertz in width, 
        the spectra from standard observations can be cleaned (baseline removal) separately and then
        co-added directly when the lines are narrow enough (a few \kmps), instead of performing
        the traditional ON minus OFF data reduction.   This technique works for
        frequency-switched observations as well as for position- and beam-switched
        observations when the ON and OFF data are saved separately.}} 
   {{ The method works best when the lines are narrow enough 
   and not too numerous so that a secure baseline removal can be achieved}. 
   A signal-to-noise ratio improvement of a factor of  $\sqrt{2}$  is  found{  in most cases, consistent
   with theoretical expectations}.}
   {By keeping the traditional observing mode, the fallback solution of the standard reduction technique is still available in cases of
   suboptimal {base}line behavior, sky instability, or wide lines, {and to confirm the line }intensities. These techniques of total-power-mode reduction can be applied to any radio
    telescope with stable baselines as long as they record {and deliver } the ONs and OFFs separately, as is the case for the GBT.}



   \maketitle
%

\section{Introduction}
For a long time, heterodyne receivers in the millimeter (mm) domain were made of front-end mixers {with optical injection of the Local Oscillator}
 followed by intermediate-frequency amplifiers and production of spectra via filter banks for  which the homogeneity of the gain was not secured from channel to channel.
Combined to on-axis telescopes, these suffer from standing waves between the secondary mirror and the receiver.  Local Oscillator injections
via Martin-Pupplet optical diplexer devices with relatively narrow bandpass were adding to the {frequency-dependent gain variations} of the whole system (a few telescope 
descriptions can be found in, e.g., \citealt{Castets1988,1989A&A...216..315B,2004A&A...423.1171S}).
{Combining the effects from the {frequency response} of the mixer, the noisy
receivers, the inhomogeneous backend,  and the intermediate-frequency amplifiers {with their imperfect impedance match to the mixer output} resulted in complex
bandpass profiles with high amplitudes, making the weak astronomical lines
difficult to detect without subtracting a reference spectrum with a
similar profile to the observations. }This subtraction has to be performed relatively quickly to protect against gain variations
and atmospheric absorption fluctuations. Various strategies are employed to get a reference spectrum that is as close as possible to the scientific spectrum. 
{These include the well-known position-switching (the telescope primary
mirror is shifted to an emission-free region in the sky), beam-switching
(the beam is deviated from the telescope pointing direction by a mirror or
simply by wobbling the secondary mirror within a few arcminutes away from
the pointed direction), and frequency-switching observing modes (where
there is no mechanical movement, but the Local Oscillator reference
frequency is slightly changed so that the subtraction does not cancel the
line but only the baseline offset).}

{For} the Rayleigh-Jeans temperature scale, the noise fluctuation (root mean square deviation, $\sigma$ or rms hereafter) of a radio spectrum is given by the radiometer 
equation:
\begin{equation}\label{eq:radiometer}
\sigma( T) = \frac{\eta T_{sys}}{\sqrt{\delta\nu\times\tau}}
,\end{equation} 

where $T_{sys}$ is the {total system noise temperature\footnote{{For the GBT, T$_{sys}$ is measured in the T$_a$ scale, while for the IRAM 30-m telescope, 
it is measured in the T$_a^*$ scale, which includes atmospheric attenuation and forward beam efficiency corrections: T$_a^*$ = T$_a\times \frac{e{^{tau}}}{\eta_{ffs}}$ 
\citep{Kutner:1981bk}}}, including the receiver
noise and all noise from the sky and ground spillover}, $\delta\nu$ is the spectral resolution, and $\tau$ the integration
time. Here, $\eta$ is sometimes introduced to take into account other losses such as one- or two-bit autocorrelator conversion losses. When using beam-switching or position-switching 
modes, a second spectrum of similar characteristics is obtained to be subtracted from the first
one. Because the noise fluctuations of both spectra are uncorrelated, the subtraction increases the noise fluctuation temperature by $\sqrt{2}$.
If we consider that $\tau$ represents the total (ON+OFF) time, then the noise fluctuation temperature increases by another $\sqrt{2}$, as only half of the time
was spent on each individual spectrum, effectively doubling the
final noise fluctuation of the spectrum (and more time is lost {in overheads, like}  the mechanical displacement of mirrors, or the rotation of the whole telescope). 
If the frequency-switch mode has been used instead, then the OFF spectrum becomes a second ON
spectrum and the penalty is only $\sqrt{2}$ { but at the expense of spoiling the baseline because of the frequency dependence 
of the receiver (and the standing wave patterns) that prevents an exact
superposition of the baseline structure from the ON1 and ON2 spectra.}
{Improved strategies consist of integrating over longer periods of time for the OFF spectrum and
subtracting the same OFF spectrum} from several ONs, minimizing the time spent OFF source
and the noise fluctuations added in the subtraction. These strategies require very stable receivers and sky conditions to be efficient, but can marginally beat
the performance of the frequency-switch mode in terms of noise fluctuations{ and provide }much flatter baselines; however, they become less efficient in the end if spatial smoothing is {applied, because the OFF is identical for the adjacent ON-OFF pairs such
that their average does not diminish the noise fluctuations efficiently,
and then frequency-switching becomes preferable}.

With the advent of high-frequency amplifiers, drop of LO optical diplexer injection, Fourier-transform spectrometers, 
and the benefit of an off-axis telescope such as the Green Bank Observatory 100-m 
Telescope (GBT), the quality and stability of the bandpass has considerably improved, meaning that total power (staring in a fixed direction  at 
fixed frequency) mode observations can now be reconsidered.
 This would avoid the quadratic addition of noise fluctuations and therefore save the $\sqrt{2}$ factor discussed above. 
 However, there are caveats in operating in pure total power mode, and we propose that observations be run in the usual way, but
with the data reduction performed on individual ON and OFF spectra to exploit the benefit of total power mode when possible while not losing
the fallback possibility of standard data reduction. We present the method in Sect. 2 and discuss a few cases in Sect. 3.

\section{Method}
The calibration of mm-wave radiotelescopes has been discussed in \citet{Kutner:1981bk} and a technical report from the 
Institut de Radio-Astronomie Millim\'etrique (IRAM) 30-m presents the details of the calibration 
procedure\footnote{\label{Kramer}\url{http://www.iram.es/IRAMES/mainWiki/CalibrationPapers?action=AttachFile\&do=view\&target=kramer\_1997\_cali\_rep.pdf} 
}
 for that telescope which is representative of mm-wave radiotelescope calibrations.
 {Nevertheless, there are differences between the GBT and the IRAM 30-m telescope. For example, the GBT
 has amplifiers before the heterodyne mixer and has no image sideband.}
 
 Briefly, the observations of two internal loads at different temperatures give a linear fit between backend voltages or detector counts and 
 temperatures (in the Rayleigh-Jeans approximation\footnote{{The physical temperature of the loads is usually 
 considered to be equal to their Rayleigh-Jeans radiation temperature. This is only true for $h\nu << kT$, with $h$ and $k$ being the Heisenberg and Boltzmann constants, $\nu$ the 
 frequency, and $T$ the physical temperature. See footnote \ref{Kramer}, and Sect. 1.3 of that paper for further considerations on the validity of the approximation.}}). The slope of the fit is referred to as the 
 gain and 
 is defined by
 \begin{equation}\label{eq:gain}
 g = \frac{T_{hot} - T_{cold}}{V_{hot} - V_{cold}}
 ,\end{equation}

where $T_{hot}$ (resp. $_{cold}$) represents the temperature of the hot (resp. cold) load and $V_{hot}$ (resp. $_{cold}$) represents the voltage (or counts) of the hot (resp. cold) load 
measured at the receiver backend. If $T_{rec}$ is the receiver equivalent noise temperature {(its value is derived from the gain measurement)}, any signal from the sky delivers 
a voltage 
$V_{sky}$ which can
be converted to a Rayleigh-Jeans temperature with
\begin{equation}\label{eq:tsky}
T_{sky} = g\times V_{sky}-T_{rec}
.\end{equation}

The sky signal is a composite of ground-emission spillover, atmospheric emission, and cosmic signal attenuated by the atmospheric absorption
and is beyond the scope of this discussion (see \citealt{Kutner:1981bk} for more information). 
If two measurements of the sky are performed along one of the switching procedures described in Sect. 1, we can compute their
difference to cancel all unwanted signals and retrieve only the cosmic signal of interest  ($T_{sky}$ is hereafter referred to with the
more commonly used $T_a$ or antenna temperature{, and similarly $V_{sky}$ becomes $V_a$}):

\begin{equation}\label{eq:deltaT}
\Delta T_a = T_{a,ON}-T_{a,OFF} = g\times(V_{a,ON}-V_{a,OFF})
,\end{equation}

where $\Delta T_a$ represents the cosmic signal we want to observe (yet uncorrected for various losses).
If we define
\begin{equation}\label{eq:tsys}
T_{sys} = T_{a} + T_{rec}
,\end{equation}
we get 
\begin{equation}\label{eq:tsysgain}
T_{sys} = g\times V_a
\end{equation}
and eq. \ref{eq:deltaT} {becomes} the more familiar form 
\begin{equation}\label{eq:deltaTsys}
\Delta T_a =  T_{sys}\times\frac{(V_{a,ON}-V_{a,OFF})}{V_{a,OFF}}
.\end{equation}

Here, \noindent$\Delta T_a$ suffers from various sources of noise which we can analyse:

\begin{equation}\label{eq:error}
\begin{aligned}
\sigma^2(\Delta T_a) = \sigma^2(g)\times(V_{a,ON}-V_{a,OFF})^2\\
+g^2\times(\sigma^2(V_{a,ON})+\sigma^2(V_{a,OFF}))
\end{aligned}
,\end{equation}
where $\sigma$ is the root mean square error expressed in a similar form as in eq. \ref{eq:radiometer}, that is, depending
on the inverse square root of the integration time and frequency sampling of the radiometer. Where the ON and OFF signals do not cancel (i.e. inside 
the observed line), the equation can be 
expressed in the familiar form, 
\begin{equation}\label{eq:errorlog}
\frac{\sigma^2(\Delta T_a)}{\Delta T_a^2} = \frac{\sigma^2(g)}{g^2}+\frac{\sigma^2(V_{a,ON})+\sigma^2(V_{a,OFF})}{(V_{a,ON}-V_{a,OFF})^2}
.\end{equation}
In the rest of the spectrum (which is usually named the baseline), the noise is simply\footnote{for a long time, mm heterodyne receivers were much noisier than the strength 
of the line and the total noise inside the line was dominated by the receiver noise. This is no longer the case for the strongest lines with modern receivers and this should be taken 
into account; though generally, the total noise becomes negligible in such cases.}
\begin{equation}\label{eq:errorbase}
\sigma^2(\Delta T_a) = g^2\times(\sigma^2(V_{a,ON})+\sigma^2(V_{a,OFF}))
.\end{equation}
If we suppose that the hot and cold load temperatures are known with enough accuracy to have a negligible 
contribution to the noise, the gain noise can be expressed as
\begin{equation}\label{eq:errorgain}
\frac{\sigma^2(g)}{g^2} = ({T_{hot} - T_{cold}})\frac{\sigma^2(V_{hot})+\sigma^2(V_{cold})}{(V_{hot} - V_{cold})^2}
.\end{equation}

If instead of observing with a switching mode (SM) one observes in total power mode (TP), the signal is not extracted
from the receiver plus sky noise during the observations; its expression is simply eq. \ref{eq:tsky}
and the error budget is (supposing $T_{rec}$ is known with enough precision to neglect its contribution)
\begin{equation}\label{eq:errorTP}
\frac{\sigma^2(T_a)}{T_a^2} = \frac{\sigma^2(g)}{g^2}+\frac{\sigma^2(V_a)}{V_a^2}
.\end{equation}

Equations \ref{eq:errorlog} and \ref{eq:errorTP} can be compared only if we have manually subtracted the sky plus receiver noise contribution
from the TP spectrum (this will be discussed in the following section), in which case $T_a(\mathrm{TP}) = \Delta T_a(\mathrm{SM})$. and $V_a(\mathrm{TP}) = 
(V_{a,ON}-V_{a,OFF})(\mathrm{SM})$.
When the receiver and the sky transparency are constant inside the bandpass of an observation, the gain and receiver noise temperature
can be calculated by integrating over all the channels of the spectrometer, largely decreasing --- suppressing in practice--- their contribution to the final noise.
The noise then simply depends on $\sigma(V_a)$, which is inversely proportional to the square root of the integration time, everything else
being equal. 
Using the same total integration time $\tau$ for the TP observations and for the sum of the two phases of the SM observations, we get

\begin{equation}\label{eq:gainbruit}
\frac{\sigma^2(\Delta T_a) }{\sigma^2(T_a)} = \frac{\sigma^2(V_{a,ON})+\sigma^2(V_{a,OFF})}{\sigma^2(V_a)} 
= \frac{{\frac{2}{\tau}}+{\frac{2}{\tau}}}{{\frac{1}{\tau}}} = 4
.\end{equation}

The SM observations are therefore twice as noisy as the TP mode ones. This is true for the
mechanical {(position or beam)} SMs because only one phase is exposed to the signal, but in the case of the frequency SM, the subsequent
folding of the spectrum reduces the noise further by $\sqrt{2}$ by averaging two independent 
realizations of the measurement and the TP mode advantage is only $\sqrt{2}$.

It is not always possible to use a constant gain and $T_{rec}$ throughout the spectrum and the balance between the gain noise and 
the observation noise contributions must be addressed when the calibration is performed channel-wise\footnote{At the IRAM 30-m the calibration scheme uses an intermediate 
configuration where the variations of the receiver temperature and sky opacity across the band are averaged by chunks of 20 MHz with their new calibration program MRTCAL.
https://www.iram-institute.org/medias/uploads/mrtcal-check.pdf}.
 We have seen (Eqs. \ref{eq:errorlog} and \ref{eq:errorgain}) that both sources of noise
depend primarily on the voltage or count measurement noise which depends upon integration time (eq. \ref{eq:radiometer}).
Since the calibrators are usually observed on a short timescale (1 -- 5 seconds typically), while the sky observations can last 60 seconds or more\footnote{we do not discuss on-the-fly 
observing mode here}, the rms of the calibration phases is the highest. In TP observations, the denominator is high 
($V_a = V_{rec}$+$V_{atmosphere}$+$V_{source}$) and comparable
to the load measurements ($V_{load} = V_{rec}+V_{cold}$ or $V_{hot}$).
Therefore, the  dominant error term comes from the calibration part itself.  In SM observations,
the observation denominator ($V_{a,ON}$-$V_{a,OFF}$)  is much smaller than $V_{load} $ and despite a
lower rms, this term dominates the gain and  calibration errors, which is preferable.
\begin{figure}
\centering
\includegraphics[width=\linewidth]{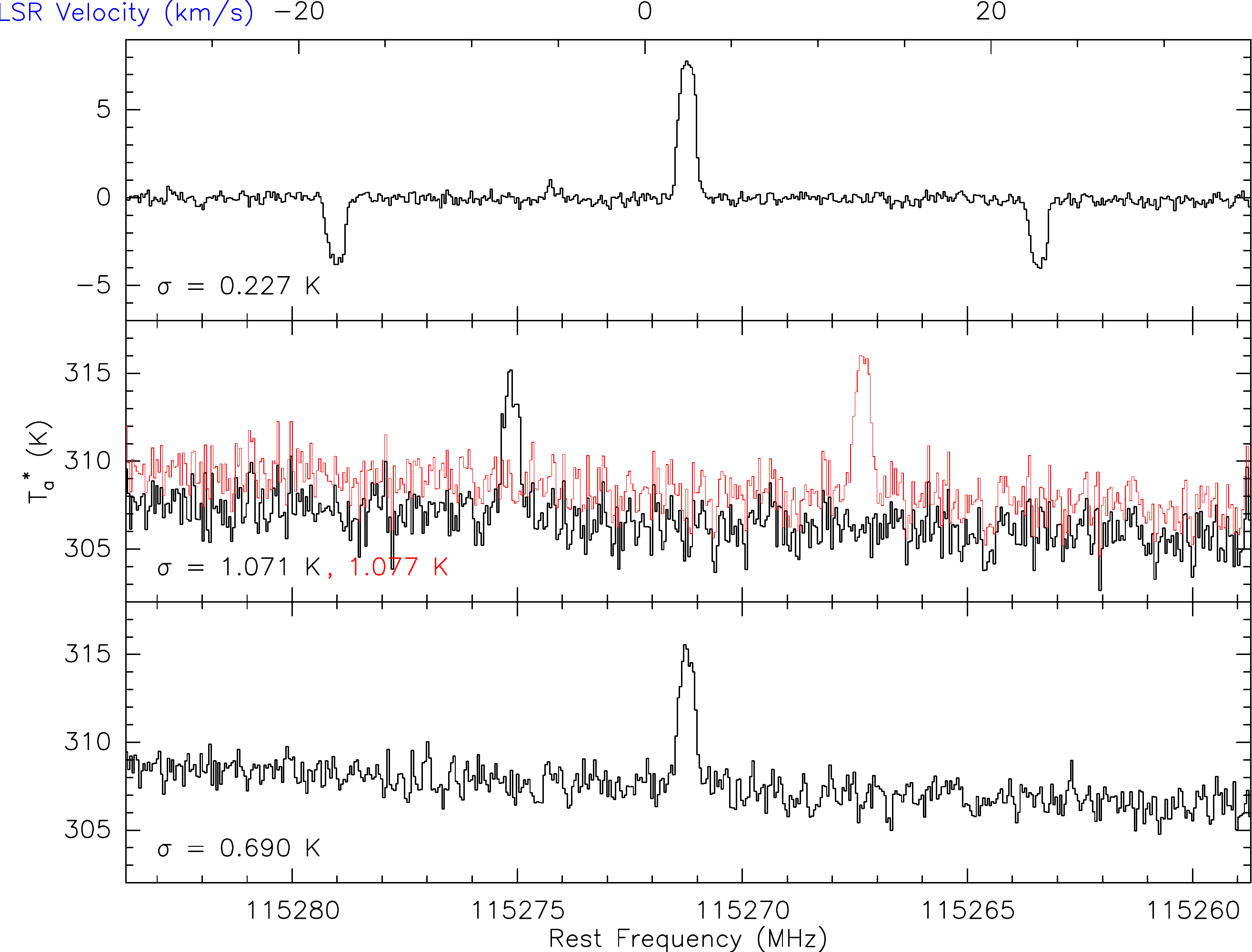}
\caption{\dzcob\juzb frequency SM observation obtained at the IRAM 30-m telescope with the EMIR receiver and FTS backend. \textbf{Top:}  Standard folded spectrum after subtraction and 
calibration. 
\textbf{Middle:} 
 Original two phases calibrated independently.      \textbf{Bottom:} Two phases directly averaged after realignment 
 {without using noise-averaged gain or subtracting a baseline from the raw data. The baseline shows no ripples but the noise is three times higher than in the upper panel}. The 
 vertical axis has the same amplitude for all
  three boxes (16 K). {The color of the rms figure corresponds to the color of the spectra to which it pertains.}}
\label{fig:calibrationfirst}
\end{figure}

Figure \ref{fig:calibrationfirst} shows an example of the problem encountered when
separately calibrating the two phases on a channel-wise mode. The final spectrum noise is dominated by the calibration noise and is therefore
noisier, which confers no advantage. To benefit from the $\sqrt{2} $ improvement, two possibilities are available:
(1) use a noise-averaged gain (either constant or smoothed over an intermediate width, as is the case for the IRAM 30-m telescope with MRTCAL), or
(2) subtract a baseline from the raw data (backend voltage or counts $V$) before applying the calibration to minimize the contribution of
the gain (calibration) noise $\sigma(g)$.

\section{Examples}

\subsection{The IRAM 30-m telescope}

Though the IRAM 30-m telescope is still equipped with front-end mixers and the antenna is on-axis, {the replacement of the optical LO injection
 by line injection} in the new EMIR receivers (for Eight MIxer Receivers, \citealt{Carter:2012dp}) has improved the baseline performance of the observations, and even 
 frequency-switched
observations are only subject to relatively limited baseline ripples. We therefore study a few cases with two different backends:
the fast Fourier Transform Spectrometer (FTS) in its narrow mode (8 $\times$ 37275 channels, 48.83 kHz each) and the VErsatile SPectrometer Array (VESPA), which is an autocorrelator used
in a narrow-window, high-resolution mode (20 MHz, 10 kHz respectively). Presently, IRAM does not deliver the individual phases (ON/OFF or Freq 1/Freq 2) of the observations
but only the final calibrated spectrum (ON - OFF or Freq 1 - Freq 2 with subsequent folding). We modified the 
calibration code (Millimeter RadioTelescope CALibration, MRTCAL) to obtain access to the 
individual raw phases for this work\footnote{{This version is incomplete and archaic and not meant to be distributed.}}.

\subsubsection{Fourier Transform Spectrometer data}
\begin{figure}
\centering
\includegraphics[width=\linewidth]{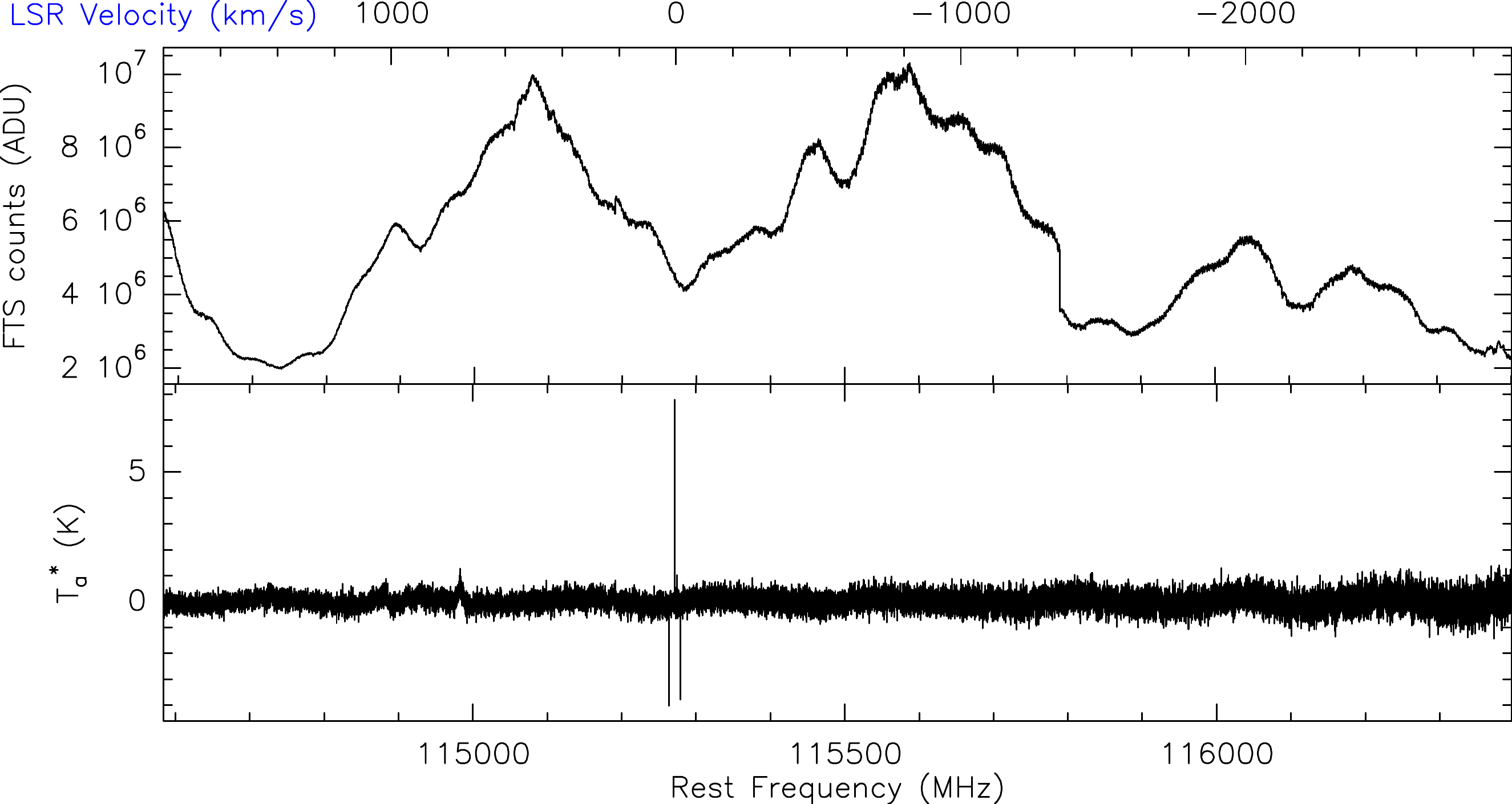}
\caption{Same observations as in Fig. \ref{fig:calibrationfirst}. \textbf{Top:} Sky observation with the full band IRAM 30-m FTS backend (one setup, one polarization).
        \textbf{Bottom:} Resulting folded frequency SM 
observation after one-minute integration. The strong positive peak with two negative half-intensity ones is the \dzcob line displayed in
the upper panel of Fig. \ref{fig:calibrationfirst}.}
\label{fig:ftstotal}
\end{figure}

Figure \ref{fig:ftstotal} clearly shows that the subtraction of a baseline from the single phase observation on the full bandwidth (upper panel) will never manage to compete with
the differential observation done in frequency SM (lower panel). There is therefore no hope to use this total power technique with wide lines and one must 
focus on a  part of the backend small enough to be able to recover a flat baseline around the line of interest.

\begin{figure}
\centering
\includegraphics[width=\linewidth]{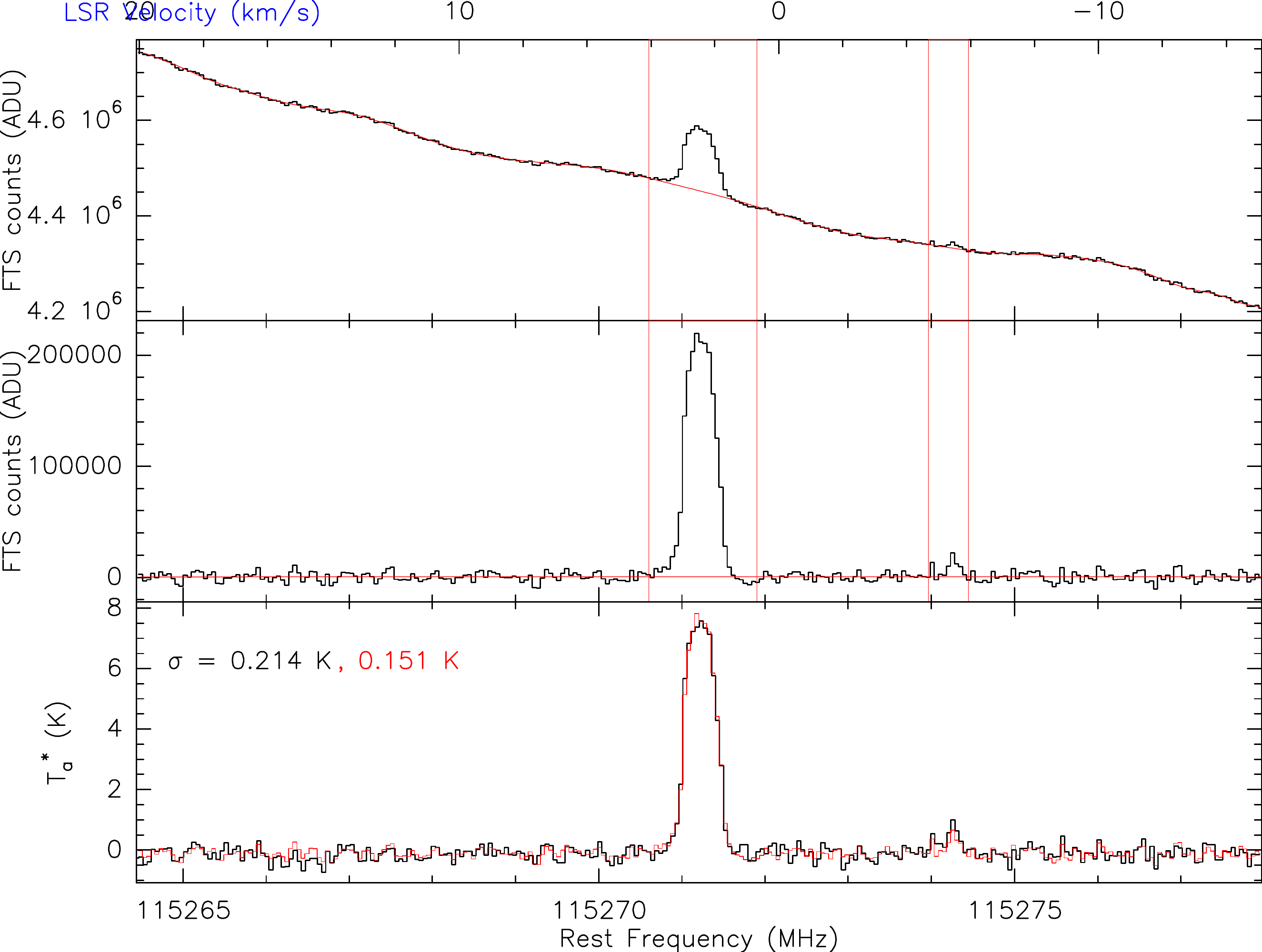}
\caption{Same observations as in Fig. \ref{fig:calibrationfirst}.       \textbf{Top:} Raw autocorrelator counts of the two phases re-aligned on top of each 
other and averaged together. A polynomial of 25$^\mathrm{th}$ order is fitted (in red). Red vertical lines {delimit} the weak telluric and strong astrophysics \dzcob lines. 
\textbf{Middle:} Baseline-subtracted raw average.       \textbf{Bottom:} Comparison between the 
standard data reduction of the observations (black) and TP mode reduction of the data (red).}
\label{fig:compareftsfswtp12co}
\end{figure}

\begin{table}[h!]
    \caption[]{Parameter optimization for TP data reduction.}
    \label{tab:parmoptim}
      \begin{tabular}{cccccc}
        \hline \hline
        \noalign{\smallskip}
        \multicolumn{3}{c}{window width}&polynomial&rms&line \\
        MHz& channels&\kmps&degree&ratio\tablefootmark{a}&ratio\tablefootmark{b}\\
        \hline
53.53 & 1096 & 139 & 40 & 0.871 & \textbf{0.998} \\
-- & -- & -- & 60 & 1.240 & 0.979 \\
-- & -- & -- & 65 & 1.275 & \textbf{1.012} \\
-- & -- & -- & 70 & 1.288 & 1.020 \\
-- & -- & -- & 75 & 1.341 & \textbf{0.998} \\
-- & -- & -- & 100 & \textbf{1.419} & 1.020 \\
        \hline
49.53 & 1014 & 129 & 55 & 1.242 & 0.980 \\
-- & -- &--& 60 & 1.274 & 1.018 \\
-- & -- &--& 65 & 1.294 & \textbf{1.008} \\
-- & -- &--& 70 & 1.341 & \textbf{0.988} \\
-- & -- &--& 90 & \textbf{1.409} & \textbf{0.995} \\
-- & -- &--& 95 & \textbf{1.411} & 0.983 \\
        \hline
45.53 & 932 & 118 & 35 & 0.929 & 0.983 \\
-- & -- &--& 55 & 1.252 & \textbf{1.010} \\
-- & -- &--& 60 & 1.267 & 1.017 \\
-- & -- &--& 65 & 1.329 & \textbf{1.000} \\
-- & -- &--& 80 & \textbf{1.389} & \textbf{0.994} \\
-- & -- &--& 95 & \textbf{1.424} & \textbf{0.988} \\
        \hline
41.53 & 851 & 108 & 30 & 0.829 & 1.020 \\
-- & -- &--& 50 & 1.251 & 1.017 \\
-- & -- &--& 55 & 1.277 & \textbf{1.006} \\
-- & -- &--& 85 & \textbf{1.408} & \textbf{1.004} \\
        \hline
37.53 & 769 & 98 & 45 & 1.236 & \textbf{1.006} \\
-- & -- &--& 50 & 1.256 & 1.016 \\
-- & -- &--& 55 & 1.332 & \textbf{0.992} \\
-- & -- &--& 75 & \textbf{1.410} & 1.016 \\
        \hline
33.53 & 687 & 87 & 25 & 0.823 & 1.021 \\
-- & -- &-- & 45 & 1.259 & \textbf{1.013} \\
-- & -- &-- & 60 & \textbf{1.379} & \textbf{1.000} \\
        \hline
29.53 & 605 & 77 & 35 & 1.218 & \textbf{1.010} \\
-- & -- &--& 40 & 1.259 & \textbf{1.013} \\
-- & -- &--& 60 & \textbf{1.409} & \textbf{1.013} \\
        \hline
25.53 & 523 & 66 & 35 & 1.316 & \textbf{0.996} \\
-- & -- &--& 45 & \textbf{1.412} & 1.019 \\
-- & -- &--& 50 & \textbf{1.433} & 1.020 \\
-- & -- &--& 60 & 1.492 & 0.985 \\
        \hline
21.53 & 441 & 56 & 25 & 1.177 & \textbf{1.007} \\
-- & -- &--& 30 & 1.276 & 0.981 \\
-- & -- &--& 40 & 1.369 & 0.984 \\
-- & -- &--& 50 & \textbf{1.424} & \textbf{1.011} \\
-- & -- &--& 65 & 1.513 & \textbf{0.987} \\
        \hline
13.53 & 277 & 35 & 15 & 1.209 & \textbf{0.992} \\
-- & -- &-- & 25 & \textbf{1.421} & \textbf{1.012} \\
-- & -- &-- & 30 & 1.473 & \textbf{1.009} \\
-- & -- &-- & 45 & 1.597 & 1.023 \\
-- & -- &-- & 50 & 1.619 & \textbf{1.008} \\
        \hline
9.53 & 195 & 25 & 10 & 1.287 & \textbf{0.994} \\
-- & -- & -- & 20 & 1.536 & \textbf{1.011} \\
         \hline
5.53 & 113 & 14 & 5 & 1.459 & \textbf{0.987} \\
-- & -- &--& 10 & 1.667 & \textbf{0.993} \\
-- & -- &--& 15 & 1.784 & 0.977 \\
 \hline
\end{tabular}    
 \tablefoot{
\tablefoottext{a}{Values close to the expected $\sqrt{2}$ improvement are boldfaced.}
\tablefoottext{b}{line ratios within 1 $\sigma$ (in terms of the integrated intensity, i.e. 0.12\,K\kmps, which is 0.986$<$R$<$1.014) are boldfaced.}
 }
\end{table}
Figure \ref{fig:compareftsfswtp12co} shows a small section of the bandpass (35 \kmps) centered on the \dzcob \juzb line. The upper panel shows the average of the two phases 
expressed in counts (analog to digital units, ADU), the cosmic and telluric line emissions are masked away 
by two windows, and a polynomial baseline of 25$^\mathrm{th}$ order has been subtracted from it. The flattened spectrum
is shown in the middle window. {The gain }to convert the spectrum to the antenna temperature scale is then applied, and the final spectrum is
compared to the original spectrum obtained by folding the two phases together. The spectra are identical and their noise is  in the
$\sqrt{2}$ ratio ($\frac{0.214}{0.151} = 1.42$). {The polynomial degree of the baseline necessary to achieve this result is a function of the width of the spectrum onto which 
the baseline is fitted.  Though this might be specific to each set of observations, we have  explored the combination of frequency window size and baseline 
polynomial degree to check their interdependence and evaluate the minimum polynomial degree needed to retrieve a consistent line-integrated intensity and a baseline flat enough to
recover the theoretical $\sqrt{2}$ improvement on the noise (Table \ref{tab:parmoptim}).  It can be noted that for polynomial degrees that are too high compared to the channel numbers, 
the noise drops even lower than expected. This indicates that the high polynomial degree induces the fit of the noise itself and removes part of 
it. {Consequently, the TP noise data reduction cannot be lead blindly but must be compared to the standard SM data reduction to check that both the noise and the line-integrated intensity achieve their expected values. In the case of no signal, the same procedure should also apply to get the correct noise improvement.}}

\subsubsection{VESPA data}

\begin{figure}
\centering
\includegraphics[width=\linewidth]{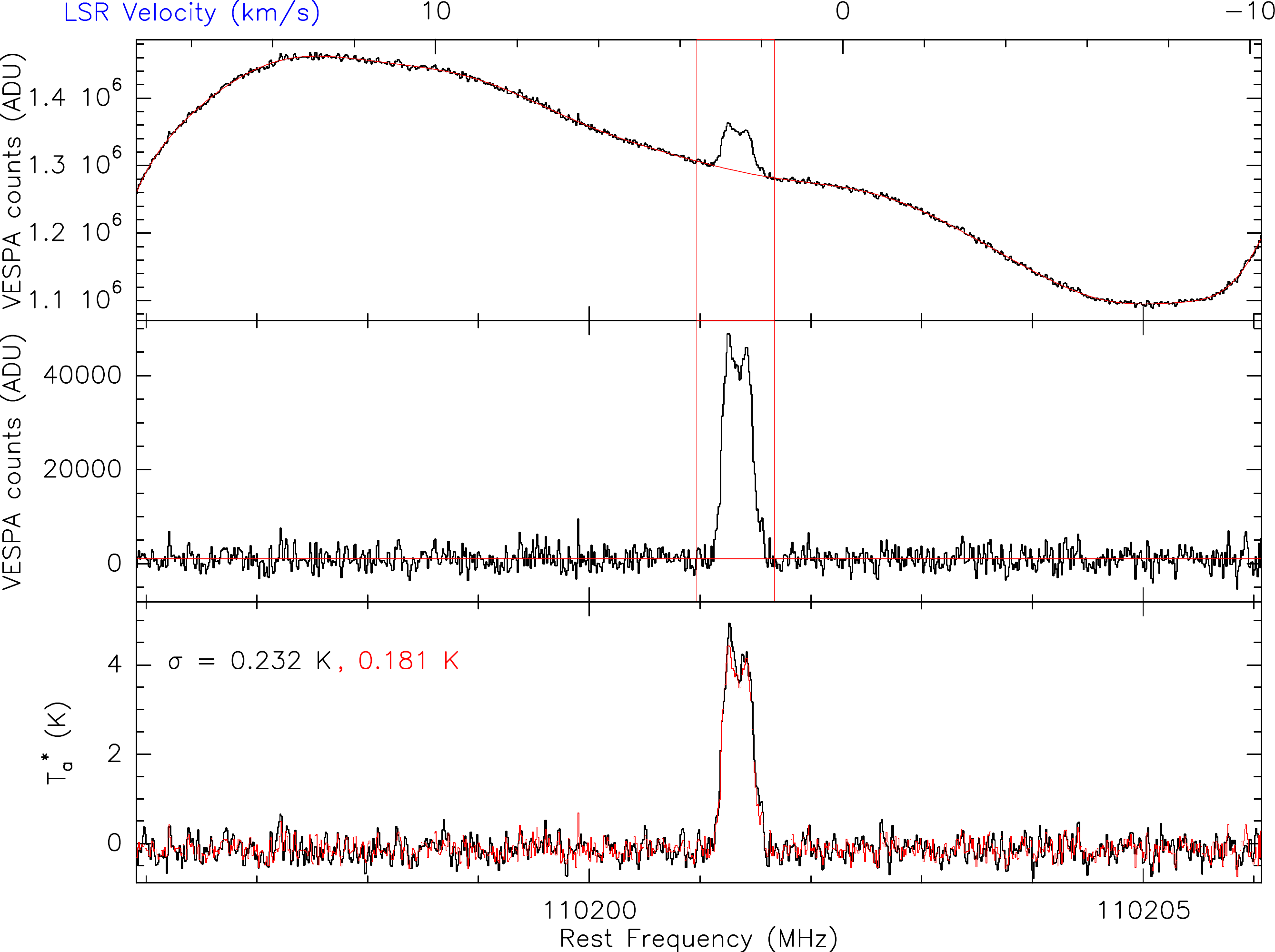}
\caption{IRAM 30-m telescope observations of the \tzcob\juzb line with the VESPA backend, full window. The original bandwidth is $\sim$20 MHz and drops to $\sim$12 MHz because of the 
$\pm$3.9 MHz 
frequency SM. The three panels display the same type of information as in Fig. \ref{fig:compareftsfswtp12co}.}
\label{fig:comparevespafswtp13coh}
\end{figure}

VESPA can be split into many small windows of 10 to 40 MHz with high frequency resolution, i.e., 3.3 to 40 kHz. There are many other
modes in addition to those discussed here, but {these are} beyond the scope of this study. On such narrow windows, the baseline is relatively smooth but still needs
polynomial fitting of high order (18-20) to remove small ripples and retrieve the correct line intensity (Fig.
\ref{fig:comparevespafswtp13coh}). However, the noise gain is less than the expected $\sqrt{2}$. We have identified the origin of
this discrepancy to the greater-than $\sqrt{2}$ noise diminution when folding the spectrum.  The measured unfolded spectrum noise
in the present case is 0.37 K, and we therefore  expect a folded spectrum noise of 0.26 K instead of the measured 0.23 K (Fig.
\ref{fig:comparevespafswtp13coh}, {with an unfolded noise of 0.33 K, expected noise of 0.23 K, and measured noise of } 0.20
K for the other polarization). Though adding or subtracting spectra should have the same impact on noise, the noise we obtain in
the TP mode data reduction is close to expected (0.37 K $\rightarrow$ 0.185 K, measured 0.181 K). The standard folding method
returns a spectrum with a noise lower than expected because the frequency displacement could be  a fractional number of
channels, and when the folding is performed the folded channels are split into two-subchannels to be added to the two adjacent channels. 
This introduces a noise correlation between the channels which can 
artificially lower the noise. We note that when half-channel intensities are added, their noise
is not changed, and therefore this is similar to smoothing the spectroscopic resolution. The diminution is maximal when the channel displacement
is exactly a whole number plus one-half of a channel and can reach $\sqrt{2}$ of supplemental noise diminution. This is not specific to VESPA data and can happen 
to any spectrometer when using the frequency SM\footnote{Observers should be aware of this artifact and should select their frequency throw to adjust to a whole number
 of channels though this might prove difficult when several backends with different resolutions are used in parallel.}. However this noise is correlated and a subsequent smoothing in 
 frequency would not reduce
the noise as much as expected. 

\subsection{The GBT 100-m}
The GBT presently uses the VErsatile GBT Astronomical Spectrometer (VEGAS)  as a backend, which is a somewhat similar  autocorrelator to VESPA in
 its agility, multi-windowing, and variable resolution
capabilities\footnote{https://www.cv.nrao.edu/~aroshi/VEGAS/skyfretobb.pdf}. Contrary to those from the IRAM 30-m telescope, the data are provided raw and uncalibrated to the user. Any 
user can therefore treat
the two phases of the observations separately in order to perform TP mode data reduction. Another autocorrelator,
known as the GBT spectrometer, was in use before VEGAS, and the results
are not sensitive to the backend itself for similar bandwidth and resolution. We visit the K-band and W-band cases here.
\subsubsection{NH$_3$ (K-band) observations }
\begin{itemize}
\item {The GBT spectrometer}
\end{itemize}
\begin{figure}
\centering
\includegraphics[width=\linewidth]{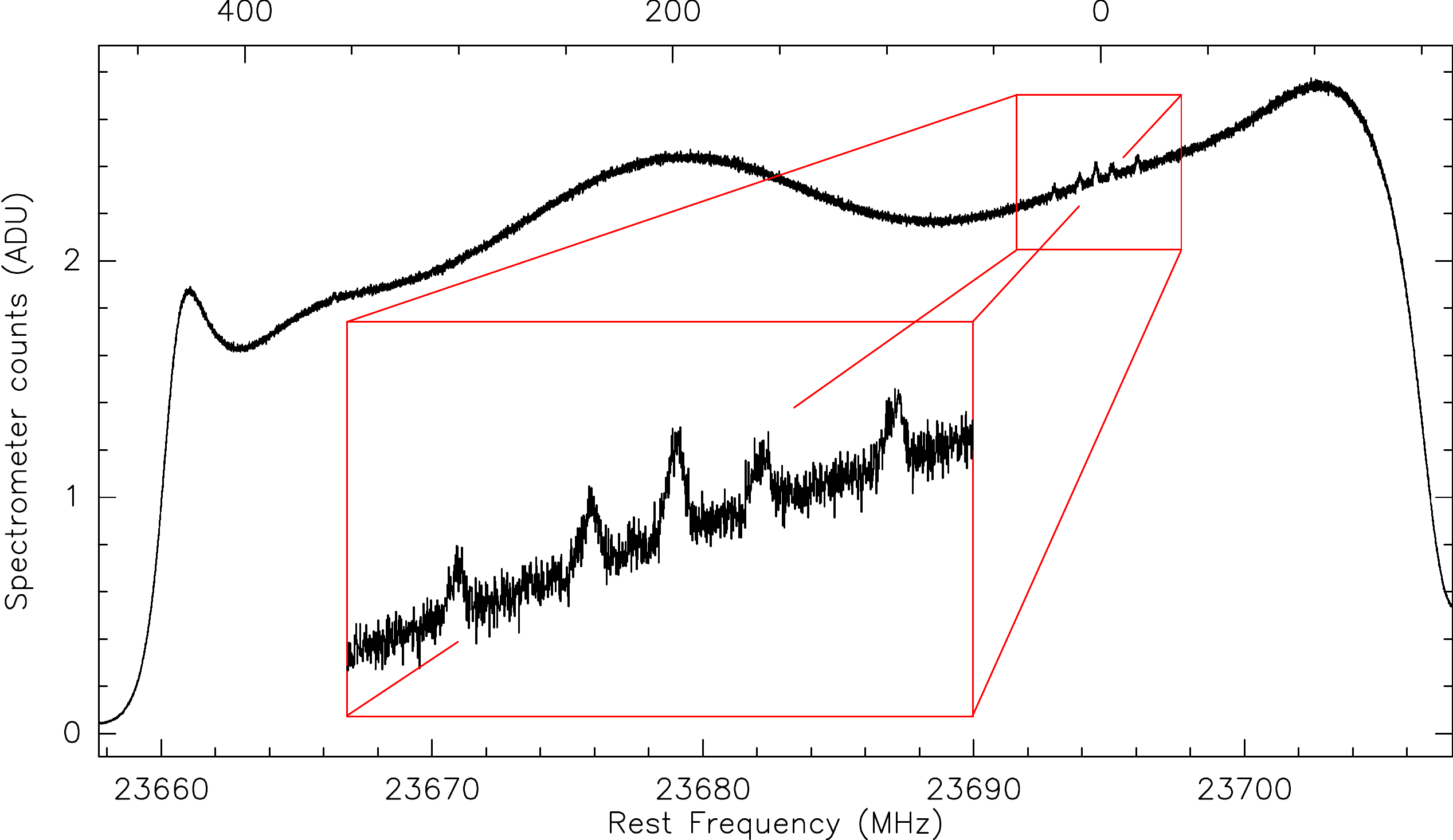}
\caption{NH$_3$ observations with the GBT 100-m in position SM. The full 50 MHz sub-window of the GBT spectrometer is represented. Only the ON phase is shown.}
\label{fig:spectrebrutnh3on}
\end{figure}

\noindent Figure \ref{fig:spectrebrutnh3on} shows the complete 50 MHz spectrum on source of a NH$_3$ line {for an }astrophysical source.
Except at the edges where the bandpass drops rapidly to zero, the band is relatively smooth and flat as for VESPA, and
the lines are readily visible in the band.

\begin{figure}
\centering
\includegraphics[width=\linewidth]{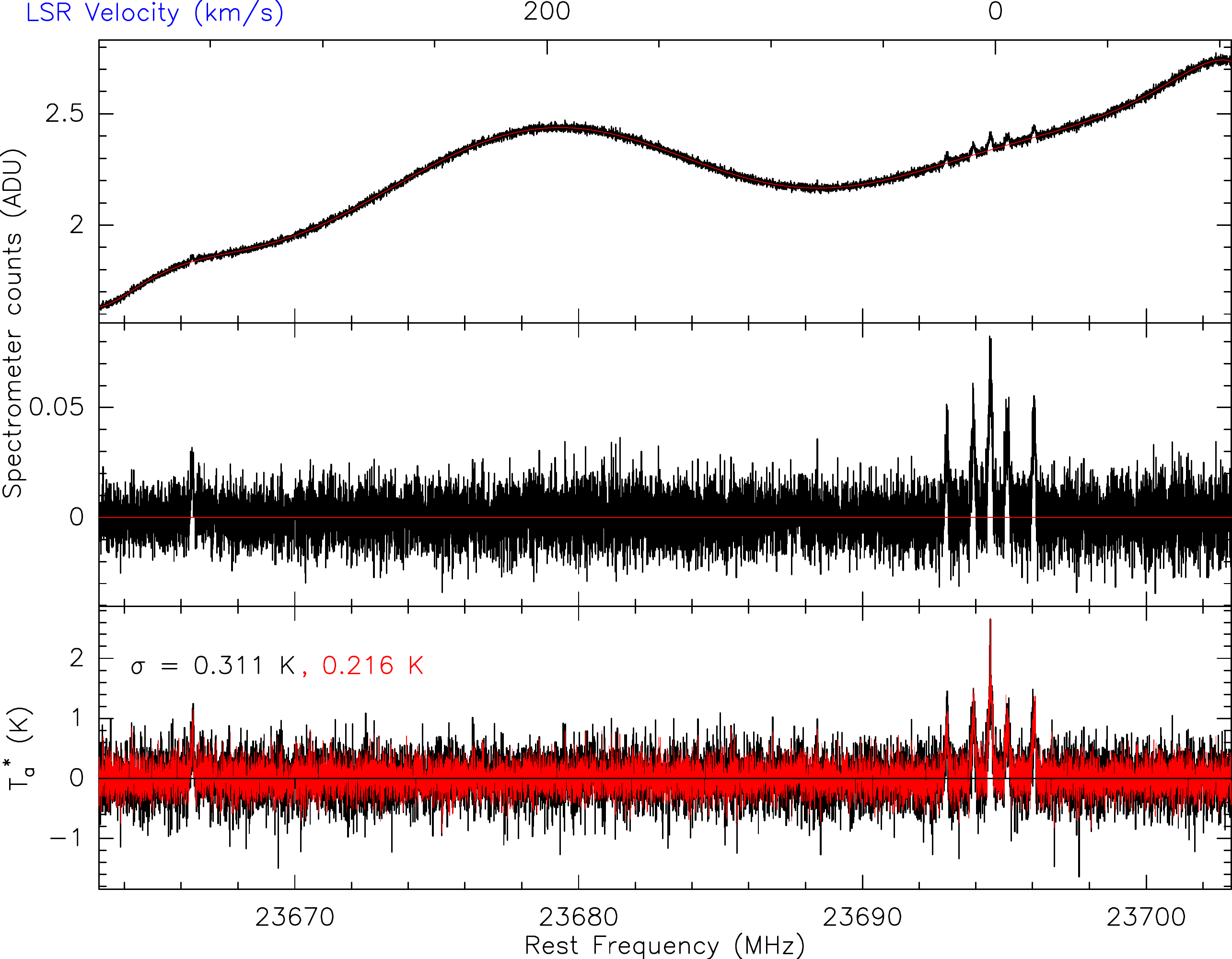}
\caption{Same observations as in Fig. \ref{fig:spectrebrutnh3on} but {dropping the ends}, with the same three types of panels as in Fig. \ref{fig:compareftsfswtp12co}}
\label{fig:comparegbtpswtpnh3}
\end{figure}

\begin{figure}
\centering
\includegraphics[width=\linewidth]{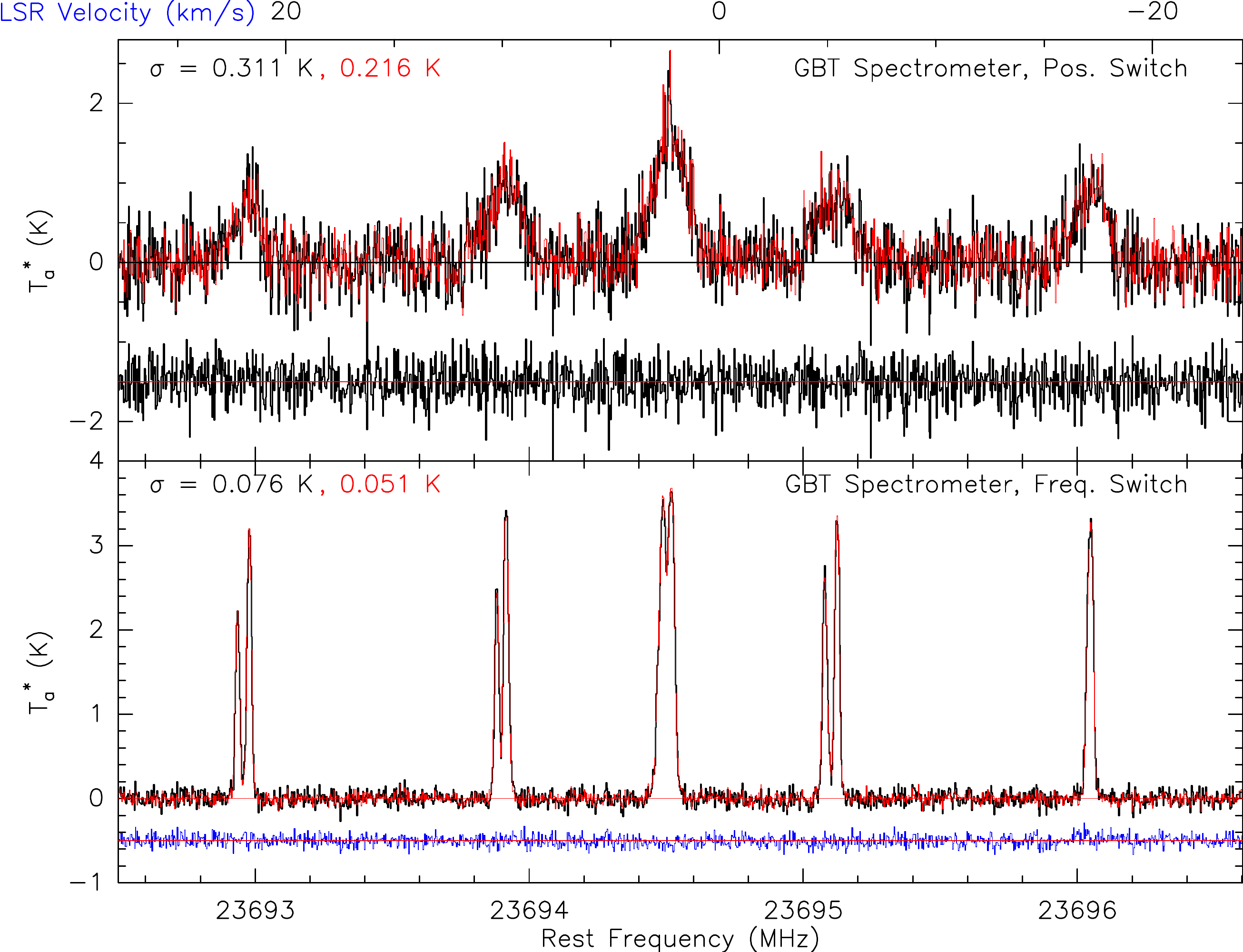}
\caption{Comparison between standard reduction (position and frequency SM, black) with TP mode reduction (red)
 for NH$_3$ observations  with the  old GBT spectrometer  in two different sources. The spectrum shown at the top 
is identical to the one displayed in Fig. \ref{fig:comparegbtpswtpnh3} but zoomed in on the lines. Noise improvement is close to $\sqrt{2}$. {The difference }between both 
reduction modes is displayed {in blue and shifted by -0.5K}.}
\label{fig:compareswtpnh32sources}
\end{figure}

A polynomial fit of 13th degree is sufficient here to remove the baseline cleanly as seen in Fig. \ref{fig:comparegbtpswtpnh3} which proves 
that in such a case, the TP mode can provide baselines as flat as those obtained naturally with mechanical SMs ({for narrow lines}).
The noise is improved by the expected $\sqrt{2}$ value, and this is also the case for the frequency SM observations of NH$_3$ with the same spectrometer (Fig. 
\ref{fig:compareswtpnh32sources}.)

\begin{itemize}
\item VEGAS
\end{itemize}
\begin{figure}
\centering
\includegraphics[width=\linewidth]{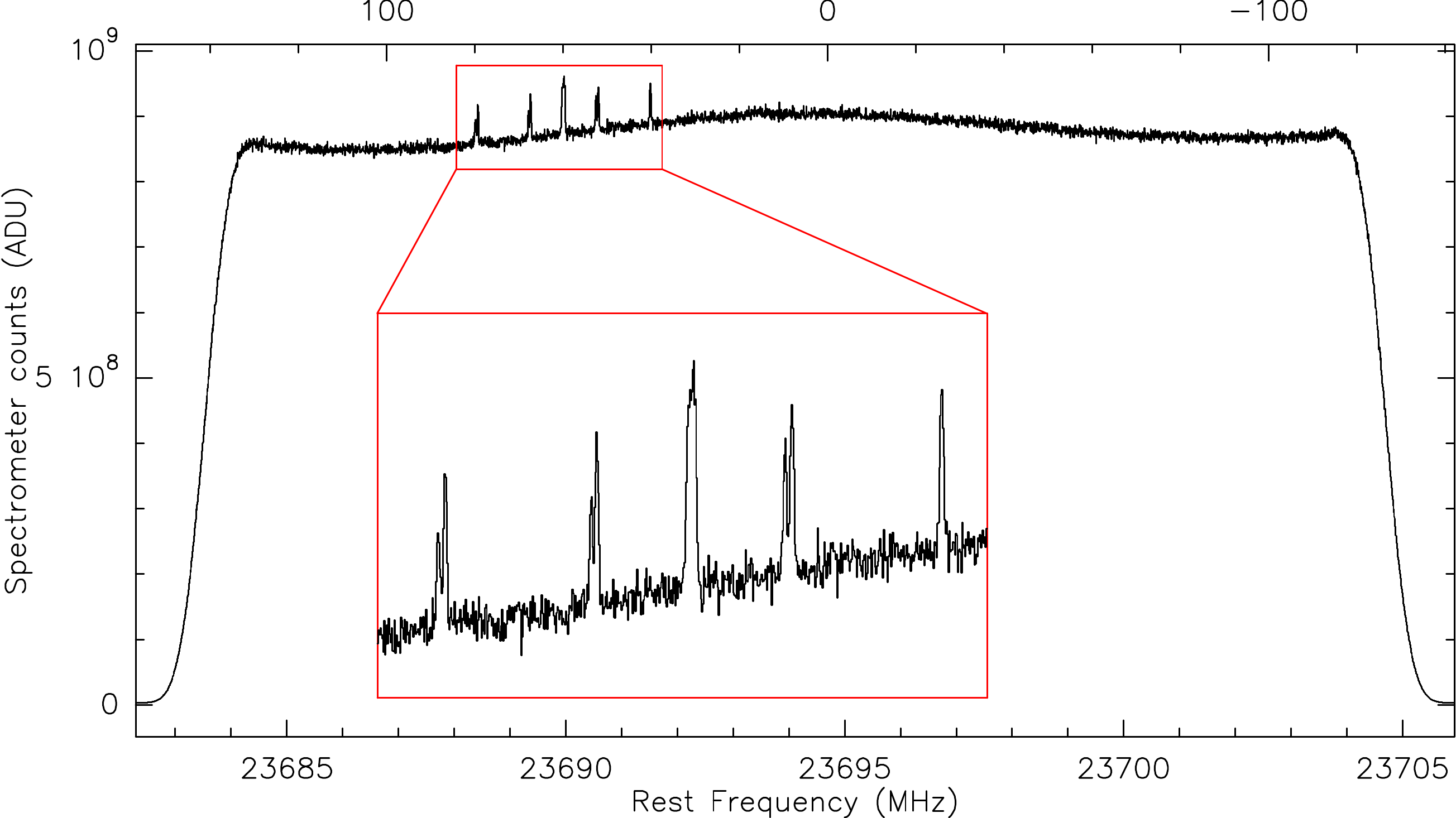}
\caption{GBT NH$_3$ observations with the VEGAS backend and the KFPA receiver, central pixel.
Only one phase of the frequency SM is shown. The full 50 MHz sub-window of VEGAS is displayed. }
\label{fig:spectrebrutnh3vegas}
\end{figure}

\begin{figure}
\centering
\includegraphics[width=\linewidth]{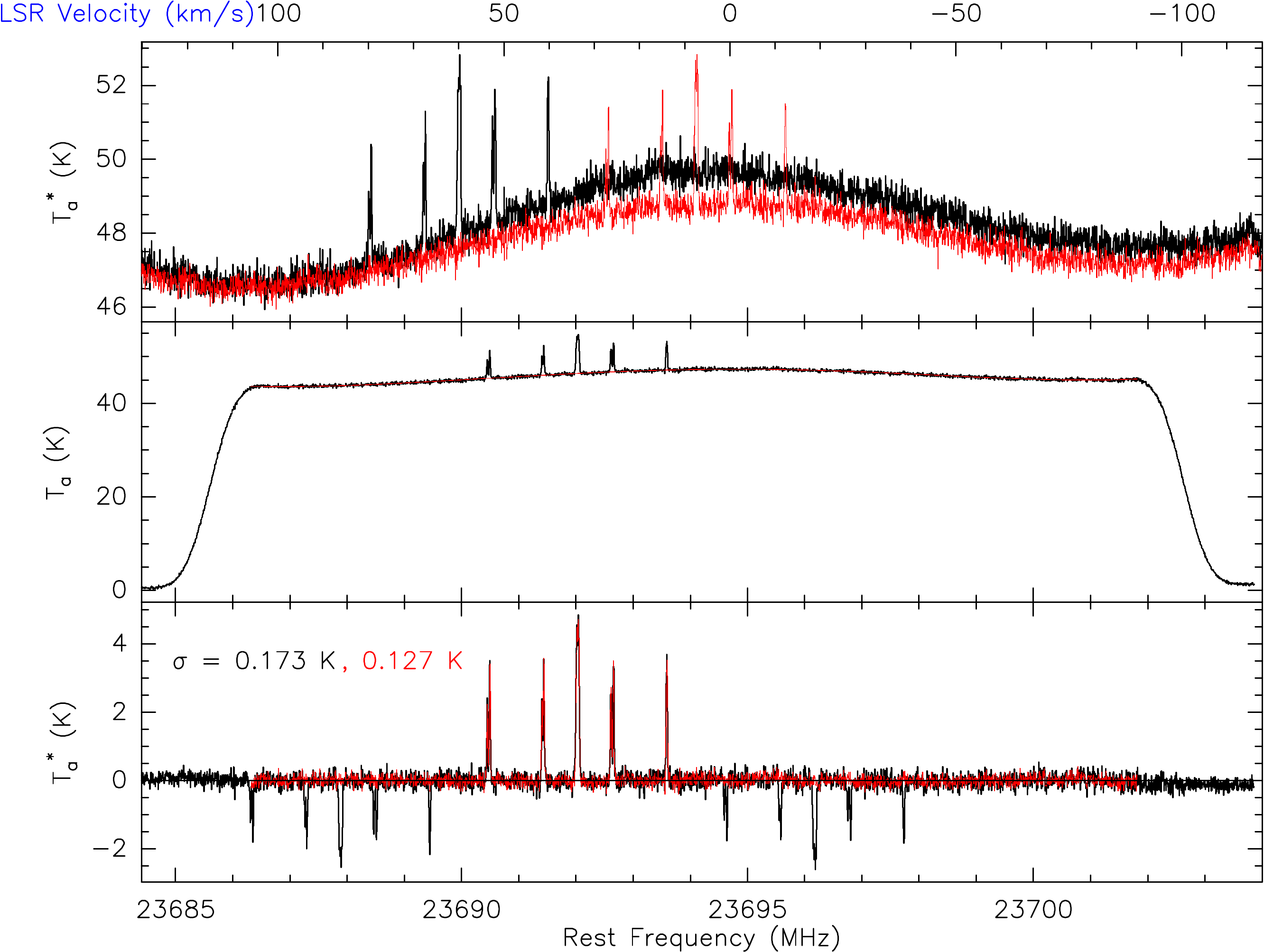}
\caption{GBT NH$_3$ observations with the VEGAS backend showing a comparison between TP mode and frequency SM. The top window shows the individual spectra at two different 
frequencies. The frequency axis is narrowed to mask the shoulders at both ends (seen in Fig. \ref{fig:spectrebrutnh3vegas}). The middle box displays the two spectra realigned and 
averaged (TP mode). Because of the 
frequency shift, both shoulders are brought back in the frequency range. A fifth degree polynomial is fitted in between them (red line). The bottom box shows the 
comparison between the standard frequency SM folded spectrum (black) and the TP mode spectrum (red, shoulders have been erased). }
\label{fig:comparespectrometervegastpnh3}
\end{figure}

VEGAS observations of the NH$_3$ line are quite similar to those provided by the GBT spectrometer, as expected, but the
baseline is even flatter, due to the new seven-pixel K-Band Focal Plane Array receiver, (KFPA, Fig. \ref{fig:spectrebrutnh3vegas}) allowing the use of a polynomial of very low 
degree (4 or 5) to zero the baseline (Fig. 
\ref{fig:comparespectrometervegastpnh3}).
The baseline is also flat enough so that the channel-wise gain is hardly variable and can be replaced by a constant value, therefore
suppressing the calibration noise from eq. \ref{eq:errorTP}. In this figure, the phases have indeed been calibrated separately before
being flattened. The use of a constant gain to calibrate the data allows the user to calibrate 
individual phases without subtracting any offset, but conversely the shoulders in the frequency SM data are hidden and their contribution to the noise lowers the average while it 
should increase it if attenuation is properly taken into account. If we measure the noise in this spectrum to the edges of the plot, which remain inside the original shoulders
 as displayed in Fig. \ref{fig:spectrebrutnh3vegas}, the standard deviation is 0.165 K instead of 0.173, 
with constant gain applied to the whole spectrum, 
 and 0.179 K if calibrated channel-wise. 

\subsubsection{W-band observations}

\begin{figure}
\centering
\includegraphics[width=\linewidth]{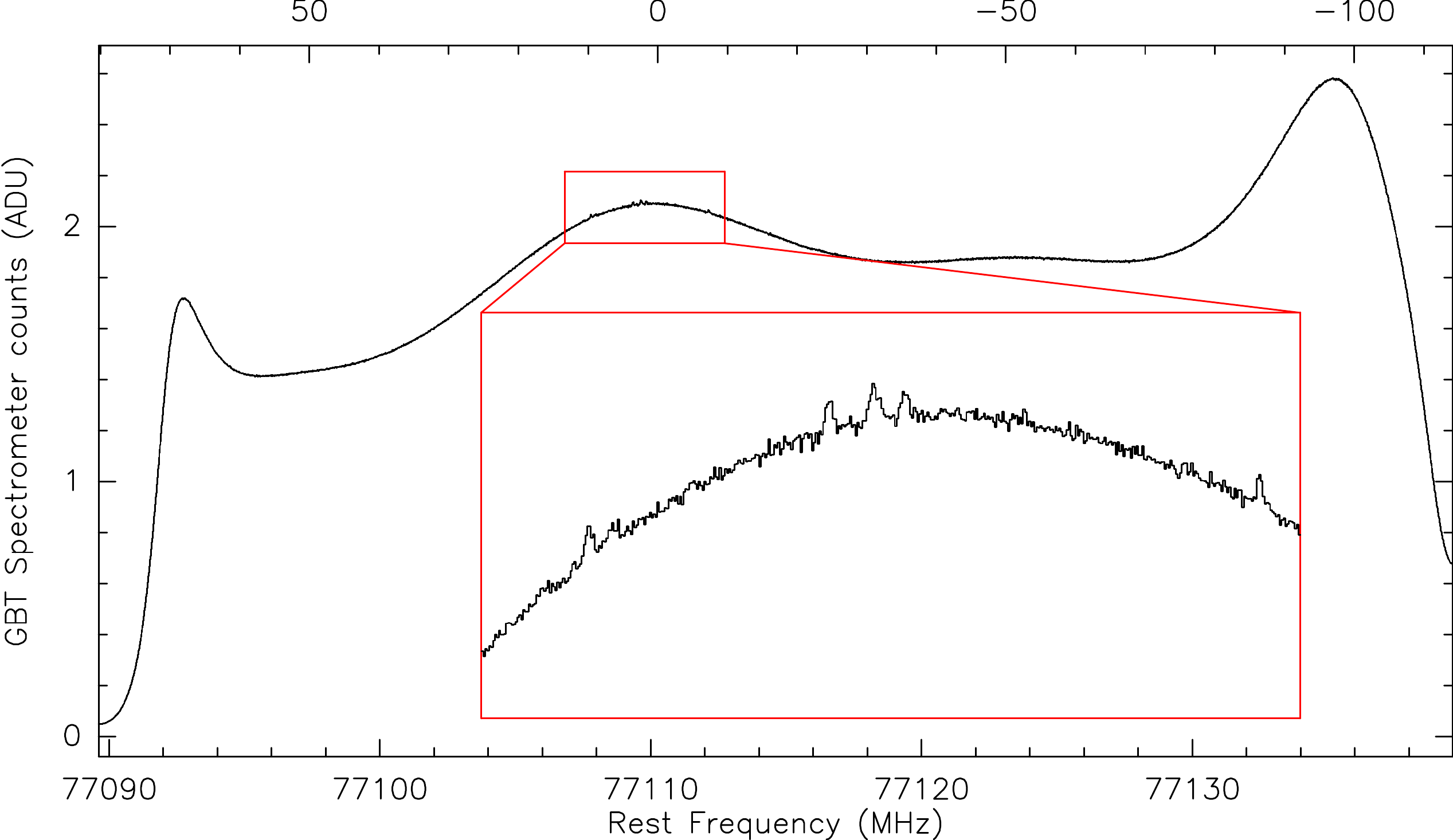}
\caption{\nddpb\juzb raw spectrum observed with the GBT spectrometer in frequency SM. The full 50 MHz bandwidth is displayed. Only one phase
is shown.}
\label{fig:spectrebrutn2dpon}
\end{figure}
\begin{figure}
\centering
\includegraphics[width=\linewidth]{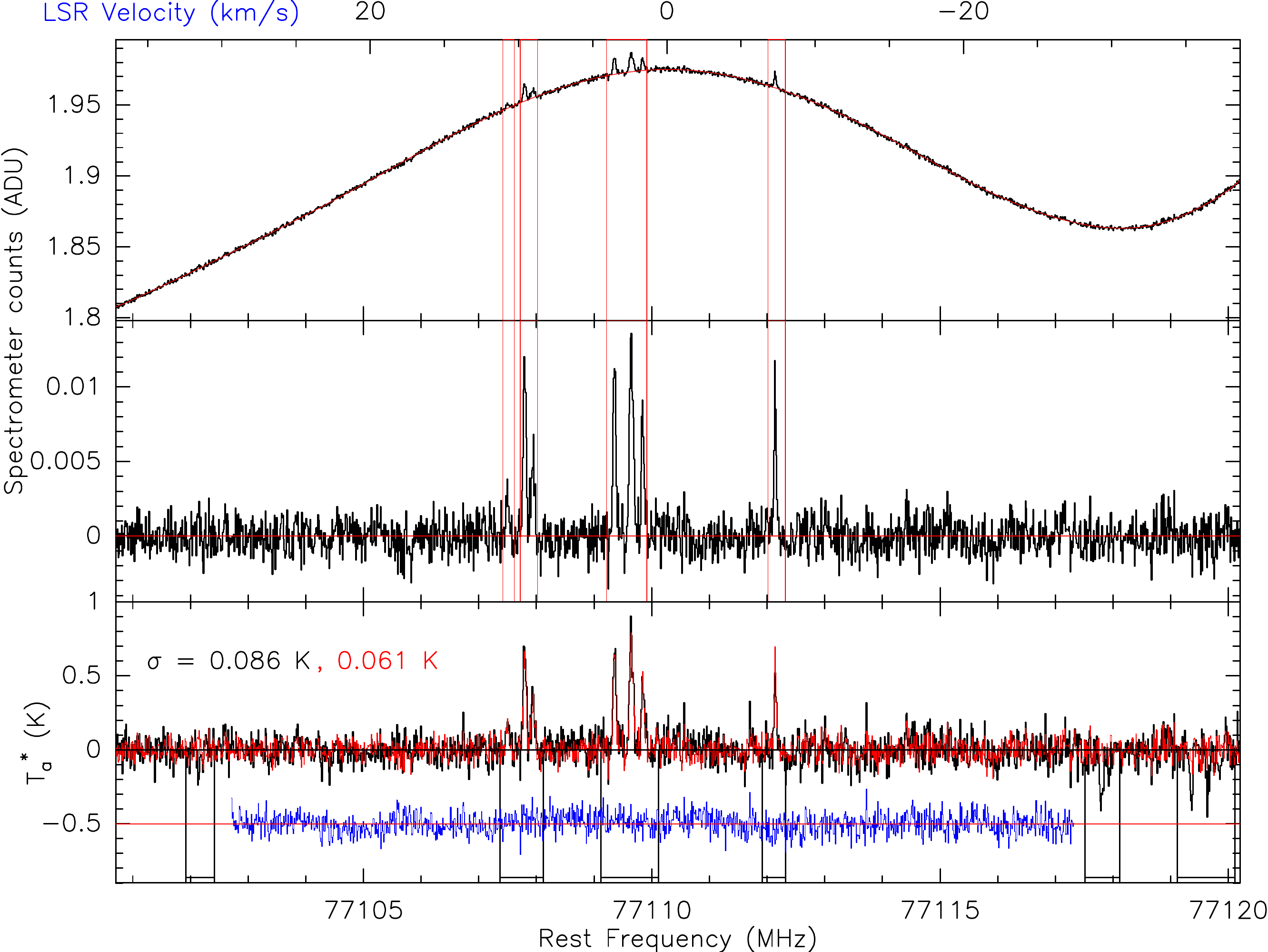}
\caption{\nddpb\juzb data reduction in TP mode compared to the original frequency SM reduction. The residual in the central part of the spectrum is shown in blue with a 
negative offset 
of -0.5 K. The ends have been erased because the negative lines from the frequency SM folding have no counterpart in the TP mode data reduction. Red vertical lines mark the 
masks used for the TP mode data reduction to compute the baseline, while the black vertical lines mark those used for the frequency SM data reduction. Part of the negative lines are
outside the window.}
\label{fig:comparespectrometerfswtpn2dpdiff}
\end{figure}
{The W-band observations we present here were taken with the old GBT spectrometer.} Though the baseline is smooth 
in appearance (Fig. \ref{fig:spectrebrutn2dpon}), the baseline subtraction from the ON+OFF spectra realigned on top of each other
 leaves small ripples which can be reduced by narrowing the spectral window but cannot be completely erased. However, the line residual is null on average, and
 the noise improvement is still $\sqrt{2}$ for a baseline polynomial degree of 16th order (Fig. \ref{fig:comparespectrometerfswtpn2dpdiff}).
 
 \subsection{Other methods}

When the line is too wide for this kind of baseline fitting, there are still various advantages to treating  the ON and the OFF observations separately.
For example, several adjacent OFF spectra in space and time (e.g., offsets of a map)
can be averaged together to be subtracted from each individual ON spectrum, reconstructing a posteriori the `one long OFF--several short ONs' observation method (which is not
available on all telescopes). The OFF can also be smoothed as strongly
as possible, as long as narrow features from the baseline itself are not smoothed out because that would replace Gaussian
noise with systematic error. 
\begin{figure}
\centering
\includegraphics[width=\linewidth]{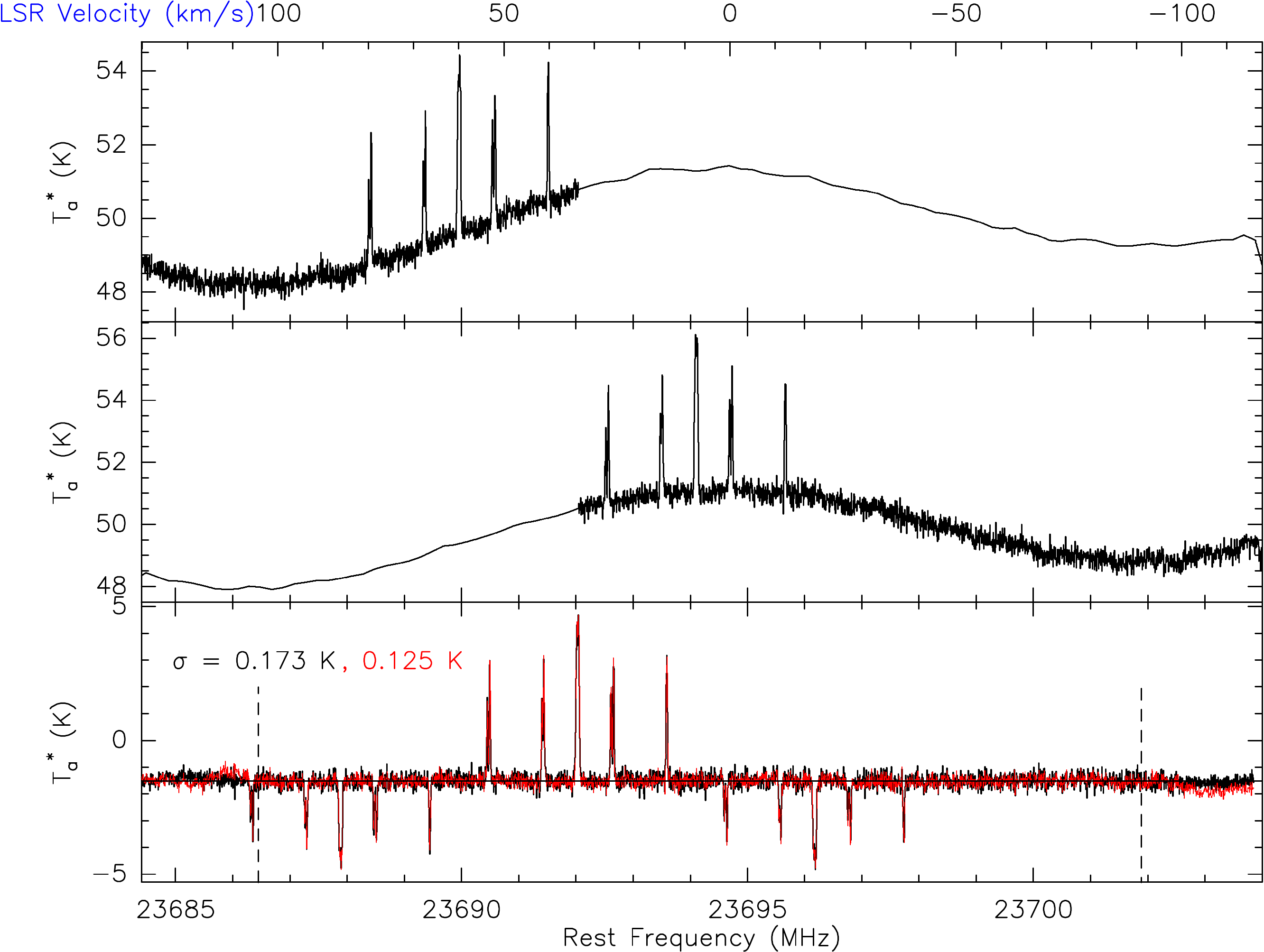}
\caption{Same NH$_3$ data from GBT + VEGAS spectrometer as in Fig. \ref{fig:comparespectrometervegastpnh3}. The top window shows the first frequency-shifted spectrum with 
the right-hand side smoothed by a median of  0.4 MHz in width. The middle window shows the second frequency-shifted spectrum smoothed on the left-hand side. The bottom window shows a 
comparison between the original folded spectrum and the half-smoothed folded spectrum. The baseline is computed between the 
two vertical dashed lines as in Fig. \ref{fig:comparespectrometervegastpnh3}.}
\label{fig:comparespectrometervegasmediannh3}
\end{figure}

For frequency SM spectra, a similar method can be applied except that the two sides of the spectra have to be treated 
separately because both contain signal. We applied this method to the VEGAS GBT NH$_3$ data with success (Fig. \ref{fig:comparespectrometervegasmediannh3})

 \section{Conclusions}
Total power data reduction with GBT data can easily be conducted with the present data-delivery and data-reduction tools (GBTIDL, a GBT
 customization of the interactive data language$\texttrademark$) offered at the GBO. 
The advantage of the present procedure is a 
gain of $\sqrt{2}$ in sensitivity at no cost whilst retaining the ability to check the result by comparing it to the normal differential data reduction or even to fall back on the
standard mode reduction in case of problems (especially if the line is too large to support a high-degree polynomial fitting  without damage). This method is particularly suited
for narrow lines in cold clouds that are usually weak and few in number per GHz. 
In principle, this method could also be used with IRAM 30-m data or any other
radiotelescope data provided the baseline is flat enough  and single phase observations are retrievable.

Though not illustrated in this paper because of a lack of suitable data, in case of wide lines, all OFF observations of different positions 
  can be averaged together (and smoothed if necessary) to reach similar results in the same manner as on-the-fly position SM
   is proceeding, and frequency SM observations can be half averaged and/or smoothed to reach sensitivities comparable to those of
   TP mode observations.
   
   Finally, OTF data can be advantageously treated in this manner: after the window width and baseline degree have been adjusted to 
   optimize the data reduction, the treatment can be applied to all spectra automatically. In particular, in the case of position SM OTF data,
   the noise contribution of the OFF spectrum does not decrease when convolving the data to produce the final map because it
   is the same OFF for all consecutive ONs. Getting rid of this noise contribution could have a noticeable impact.
\begin{acknowledgements}
We thank the anonymous referee and the editor, Thierry Forveille, for their remarks which helped to improve the clarity of the paper.
We thank S\' ebastien Bardeau for his help to modify MRTCAL and explanations on data resampling.
\end{acknowledgements}
 \bibliographystyle{aa}
 
\bibliography{/Users/laurent/bibtex/references,/Users/laurent/bibtex/papers}

\begin{thebibliography}{5}
\expandafter\ifx\csname natexlab\endcsname\relax\def\natexlab#1{#1}\fi

\bibitem[{{Booth} {et~al.}(1989){Booth}, {Delgado}, {Hagstrom}, {Johansson},
  {Murphy}, {Olberg}, {Whyborn}, {Greve}, {Hansson}, {Lindstrom}, \&
  {Rydberg}}]{1989A&A...216..315B}
{Booth}, R.~S., {Delgado}, G., {Hagstrom}, M., {et~al.} 1989, \aap, 216, 315

\bibitem[{Carter {et~al.}(2012)Carter, Lazareff, Maier, Chenu, Fontana,
  Bortolotti, Boucher, Navarrini, Blanchet, Greve, John, Kramer, Morel,
  Navarro, Pe{\~n}alver, Schuster, \& Thum}]{Carter:2012dp}
Carter, M., Lazareff, B., Maier, D., {et~al.} 2012, A{\&}A, 538, A89

\bibitem[{{Castets} {et~al.}(1988){Castets}, {Lucas}, {Lazareff}, {Cernicharo},
  {Omont}, {Duvert}, {Fouilleux}, {Forveille}, {Pagani}, {Beaudin},
  {Deschamps}, {Encrenaz}, {Lebourg}, {Gheudin}, {Perault}, {Ruffie},
  {Clavelier}, {Lacroix}, {Lauque}, {Montignac}, {Baudry}, \&
  {Champion}}]{Castets1988}
{Castets}, A., {Lucas}, R., {Lazareff}, B., {et~al.} 1988, \aap, 194, 340

\bibitem[{Kutner \& Ulich(1981)}]{Kutner:1981bk}
Kutner, M.~L. \& Ulich, B.~L. 1981, ApJ, 250, 341

\bibitem[{{Schuster} {et~al.}(2004){Schuster}, {Boucher}, {Brunswig}, {Carter},
  {Chenu}, {Foullieux}, {Greve}, {John}, {Lazareff}, {Navarro}, {Perrigouard},
  {Pollet}, {Sievers}, {Thum}, \& {Wiesemeyer}}]{2004A&A...423.1171S}
{Schuster}, K.~F., {Boucher}, C., {Brunswig}, W., {et~al.} 2004, \aap, 423,
  1171

\end{thebibliography}
 
\end{document}